\documentstyle[12pt,epsfig,aps,eqsecnum]{revtex}
\newcommand{\be}{\begin{equation}}
\newcommand{\ee}{\end{equation}}
\newcommand{\bea}{\begin{eqnarray}}
\newcommand{\eea}{\end{eqnarray}}
\newcommand{\dpf}{\displaystyle\frac}

\begin{document}
\title{Rho Meson Broadening in Hot and Dense Hadronic Matter }
\author{Song\ Gao$^1$, 
Charles\ Gale$^2$, Christoph Ernst$^1$, \\
Horst\ St\"ocker$^1$ and Walter\ Greiner$^1$}
\address{$^1$Institut f\"ur Theoretische Physik der J.W. Goethe-Universit\"at\\
Postfach 111932, D-60054 Frankfurt a. M., Germany \\
	$^2$Physics Department, McGill University\\
         3600 University St., Montr\'eal, QC, H3A 2T8, Canada}

\maketitle

\vspace*{1.0cm}

\centerline{\large Abstract}
	The modification of the width of rho mesons due to in-medium
decays
and collisions is evaluated. The decay width is calculated from
the imaginary part of the one-loop self-energy at finite temperature.
The collision width is related to the cross sections of the $\rho$ + pion
and the $\rho$ + nucleon reactions. A calculation based on
an effective Lagrangian shows the importance of including
the direct $\rho\pi\to\rho\pi$ scattering which is
dominated by the $a_1$ exchange.
A large broadening of the spectral function is found, accompanied
by a strength suppression at the pole.

\vspace*{1.0cm}


\newpage

\section{Introduction }

	It is generally believed that ultrarelativistic heavy ion
collisions produce matter far from the ground state and provide 
conditions of high
temperature and/or high density suitable for investigating modified
particle properties in the medium. The  vector mesons, e.g. the 
rho, are 
particularly important components of the matter produced in 
ultrarelativistic heavy ion collisions. It has been speculated that 
the restoration of chiral symmetry induces a mass shift of the
$\rho$ meson\cite{BRs}. The direct dilepton decay channel, 
$\rho \to \, e^+ e^-$,
appears to be an ideal probe for studying the in-medium properties 
of the $\rho$:
a possible mass shift, a modified dispersion relation, or  
a width modification \cite{hatsuda,gale91,gqli,rapp,haglin95,haglin98,klingl}.
Unfortunately, due to the interplay between those effects, it seems to be
difficult to isolate the different effects from invariant mass 
 measurement alone. 
The in-medium mass shift and broadening of the width of a particle 
have been related to the real part of the forward scattering amplitude 
on the constituents of the medium and thus to the corresponding cross
sections. These results indicate 
a small mass shift but a large increase of the width
both in intermediate and high energy heavy ion
collisions\cite{rapp,elsky,sib98}. 
The goal of the present paper is to verify these results using slightly
different techniques. We will point out the differences as we proceed.

	Recently, dilepton measurements
at the CERN/SPS energy have received quite some attention\cite{CERES}. 
The $e^+ e^-$ mass spectra (in
S + Au collision at $E/A = 200$ GeV and in Pb + Au collision at $E/A =
160$ GeV preliminary data) seem to indicate a 
$\rho$-peak suppression in central collisions. 
The absence of $\rho$-peak may be due to a large width of the $\rho$ meson
as compared with its free value. 
Of course, a confirmation of such a 
conjecture must include a dynamical simulation for
the space time evolution of the colliding system. 
In any case, it is important to
quantitatively address this issue independently. 

	The in-medium effects on the $\rho$ meson have been widely 
investigated using finite temperature field theory at the one-loop
level \cite{gale91,gsz,shi}, but the most important source for the width
modification occurs at the two-loop level \cite{haglin95}. The present
paper focuses on the $\rho$ meson width modification up to the
two-loop level. The pole shift of the
$\rho$, as emerged from many-body calculations 
is small \cite{rapp,klingl,elsky,sib98}. 

	The in-medium properties of the $\rho$ are
due to its interaction with the $constituents$ of the medium. In the
hadronic matter produced in ultrarelativistic heavy ion collisions,
pions and
nucleons are important constituents, so we take into account both
$\rho \, \pi$ and $\rho$ N reactions in our $\rho$ width calculations. 
Interactions with other hadrons are expected to increase the width above the
values found here.

	The paper is organized as follows: In section II, we present
the general formul{\ae} for the collision rate  of in-medium
 $\rho$'s. 
We present a calculation of the thermal averaged in-medium width. 
In section III, we discuss the cross sections 
for $\rho \, \pi$- 
and $\rho$N- reactions from which the collision rates are determined.
The width modifications due to $\rho \, \pi$- and $\rho$N- reactions
are calculated at finite temperature and density in section IV.
Section V concludes with a summary.

\section{Formalism}

	The principal elementary reactions which tend to 
thermalize $\rho$'s in hot and dense matter are the channels 
 $\rho_1\, \pi_2 \to \rho_3\, \pi_4$ and/or
$\rho_1\, N_2 \to \rho_3\, N_4$. A $\rho$ with arbitrary momentum $p_1$ 
(and energy $\omega$) can be captured by the thermal background. The rate 
for this direct capture \cite{weldon92} is: 
\be 
\Gamma_d (\omega) = {1 \over {2 \omega}} \int d \Omega \, n_2(E_2)\,
           (1 \pm n_3(E_3))\, (1 \pm n_4(E_4))\,
          {\overline  {|{\cal M}(12 \to 34)|^2}} \ \ ,
\label{eq:cd}
\ee

A $\rho$ with momentum $p_1$ can also be produced from the thermal background
by the inverse reaction: $\rho_3\, \pi_4 \to \rho_1\, \pi_2$. This inverse 
rate (omitted in many similar treatments) is
\be 
\Gamma_i (\omega) = {1 \over {2 \omega}} \int d \Omega \,(1 \pm n_2(E_2)) \,
            n_3(E_3)\, n_4(E_4)\,
           {\overline {|{\cal M}(34\to 12) |^2}} \ \ ,
\ee
where
\begin{equation}
d \Omega = d {\overline p_2} \, d {\overline p_3} \, d {\overline p_4} \,
           (2 \pi)^4 \, \delta(p_1+p_2-p_3-p_4) \, ,
\end{equation}
and 
$$d {\overline p_i} = {d^3 p_i \over {(2 \pi)^3\, 2 E_i}} \, .$$

$n_i(E_i)$ are Bose-Einstein or Fermi-Dirac distribution functions,  
`$\pm$' corresponds to Bosons and Fermions, respectively. 

The total collision rate for bosons is the difference 
\be
\Gamma^{\rm coll} (\omega) = \Gamma_d (\omega) - \Gamma_i (\omega)\, .
\ee  

	This expression can be derived from quantum field
theory \cite{jeon}. From finite temperature quantum field
theory, the width of any particle, 
$\Gamma(\omega)$, is given by the imaginary part of its self-energy, 
${\rm Im}\Pi(\omega)$,
\be
{\rm Im}\Pi(\omega) = - \omega \Gamma(\omega) \  ,
\ee
which provides a link between field theory and statistical mechanics. 

	The ratio of $\Gamma_d
(\omega)$ to $\Gamma_i (\omega)$ is an universal function of $\omega$
\cite{weldon83},
\be 
\dpf{\Gamma_d(\omega)}{\Gamma_i (\omega)} = {\rm exp}(\omega/T)\ , 
\ee
because the single-particle distribution functions satisfy
\bea
  \dpf{1+n_B(\omega)}{n_B(\omega)} &=& {\rm exp}(\omega/T) \ \ ( {\rm Boson} ) \ \ , \cr
  \dpf{1-n_F(\omega)}{n_F(\omega)} &=& {\rm exp}(\omega/T) \ \ ( {\rm Fermion} ) \ \ .
\eea
Thus the total collision rate is given by:
\be
\Gamma^{\rm coll} = (1 - e^{-\beta \omega}) \, \Gamma_d (\omega)\ .
\ee 

In the simplest scenario of equilibrated hot and dense matter, 
we can calculate the thermal 
average of the collision rate:
\be
{\overline \Gamma}^{\rm coll} = {{\large \int} {\rm d}^3 p_1 
	\Gamma^{\rm coll} \, n_1(\omega) \over
          {\int {\rm d}^3 p_1 \, n_1(\omega)}} =
          {\int {\rm d}^3 p_1 \Gamma_d \, e^{-\beta \omega} \over
          {\int {\rm d}^3 p_1 \, n_1(\omega)}} \, ,
\label{eq:avc}
\ee
Using (\ref{eq:cd}) and (\ref{eq:avc}), the thermal rate can be written as
\be
{\overline \Gamma}^{\rm coll} =\dpf{{\cal N}_1 {\cal N}_2}{\rho_1}\,
        \int_{s_0}^{\infty} {\rm d} s \dpf{T}{2(2\pi)^4 {\sqrt s}}\, 
	\lambda(s, m_1^2, m_2^2)\, {\tilde K}({\sqrt s}, m_1, m_2, T, \mu_i)\,
	\sigma_{12\to 34}(s) \, ,
\label{eq:aver}
\ee
where 
\be 
\rho_1 = {\cal N}_1 \int \dpf{{\rm d}^3 p_1}{(2\pi)^3}\,n_1(\omega)\, 
\ee 
is the number density of the $\rho$'s we considered,
 $\sigma_{12\to 34}(s)$ is the cross section of $\rho \, \pi$ (or $\rho
\, N$) collisions, and $s_0=(m_1+m_2)^2$, ${\cal N}_i$ are the spin-isospin
degeneracy factors.   
The function ${\tilde K}$ assumes different forms for different
statistics\cite{lich}. Below we apply classical Boltzmann statistics,
where
\be
{\tilde K} = K_1({\sqrt s}/T) {\rm exp}[(\mu_1 + \mu_2)/T] \, ,
\ee
with $K_1$ being the modified Bessel function. $\mu_1 \, , \mu_2$ are the
chemical potentials of $\rho$ and $\pi$ (or nucleon) respectively. 
 From (\ref{eq:aver}) the collision rate of $\rho$ in the hot 
and dense matter can be obtained 
if the cross sections of the $\rho\, \pi$- or 
$\rho$N- processes are known.

\section{$\rho\pi$ and $\rho$N cross sections}

	 For $\rho \, \pi$ collisions, we use both the
scattering amplitude of the elementary processes method and
the resonance contribution approximation to calculate the
cross sections $\sigma_{\rho\,\pi}$. 
We consider the
scattering amplitudes for the elementary reactions in
Fig.~1 which can be obtained by cutting the two-loop
diagrams for the $\rho$ meson self-energy as demonstrated
in Ref. \cite{kapusta91}.

To determine the $\rho$ N cross section 
$\sigma_{\rho\, N}$, a method based on the resonance model is used to 
get a lower estimate. 

\subsection{$\rho \, \pi$ cross section}

The standard Breit-Wigner formula for $\rho \, \pi$  cross section is
\be
\sigma_{\rho\,\pi} = \dpf{\pi}{q^2}\, \sum_R F_s\, F_i\,
	\dpf{B_R \Gamma_R^2}{({\sqrt s}-m_R)^2+\Gamma_R^2/4}  \ \ .
\label{eq:rhopi}
\ee
Here `R' refers to the intermediate resonance, and $B_R$ is
the branching ratio of its decay into $\pi \, \rho$. ${\sqrt s}$ is the
total center-of-mass energy, $q$ is the three-momentum of the 
$\rho$ meson in the resonance frame,
\be 
q = \dpf{1}{2{\sqrt s}}\,\lambda^{1/2}(s, m_{\pi}^2, m_{\rho}^2)\, ,
\ee
with the kinematical triangle function 
$\lambda(x,y,z) = x^2-2x(y+z)+(y-z)^2$. $F_s$ and $F_i$ are 
spin and isospin degeneracy factors:
\be 
F_i =\dpf{2I_R+1}{(2I_{\pi}+1)(2I_{\rho}+1)} \ \ \ , \ \ \ \ \ \ \
F_s = \dpf{2J_R+1}{(2s_{\pi}+1)(2s_{\rho}+1)} \, .
\ee

We take into account the following resonances: $\phi(1020)$, 
$a_1(1260)$, $a_2(1320)$, $\omega'(1420)$, $\pi(1670)$, with the  
properties listed in \cite{PRD}.  

\begin{figure}[hbt]
{\makebox{\epsfig{file=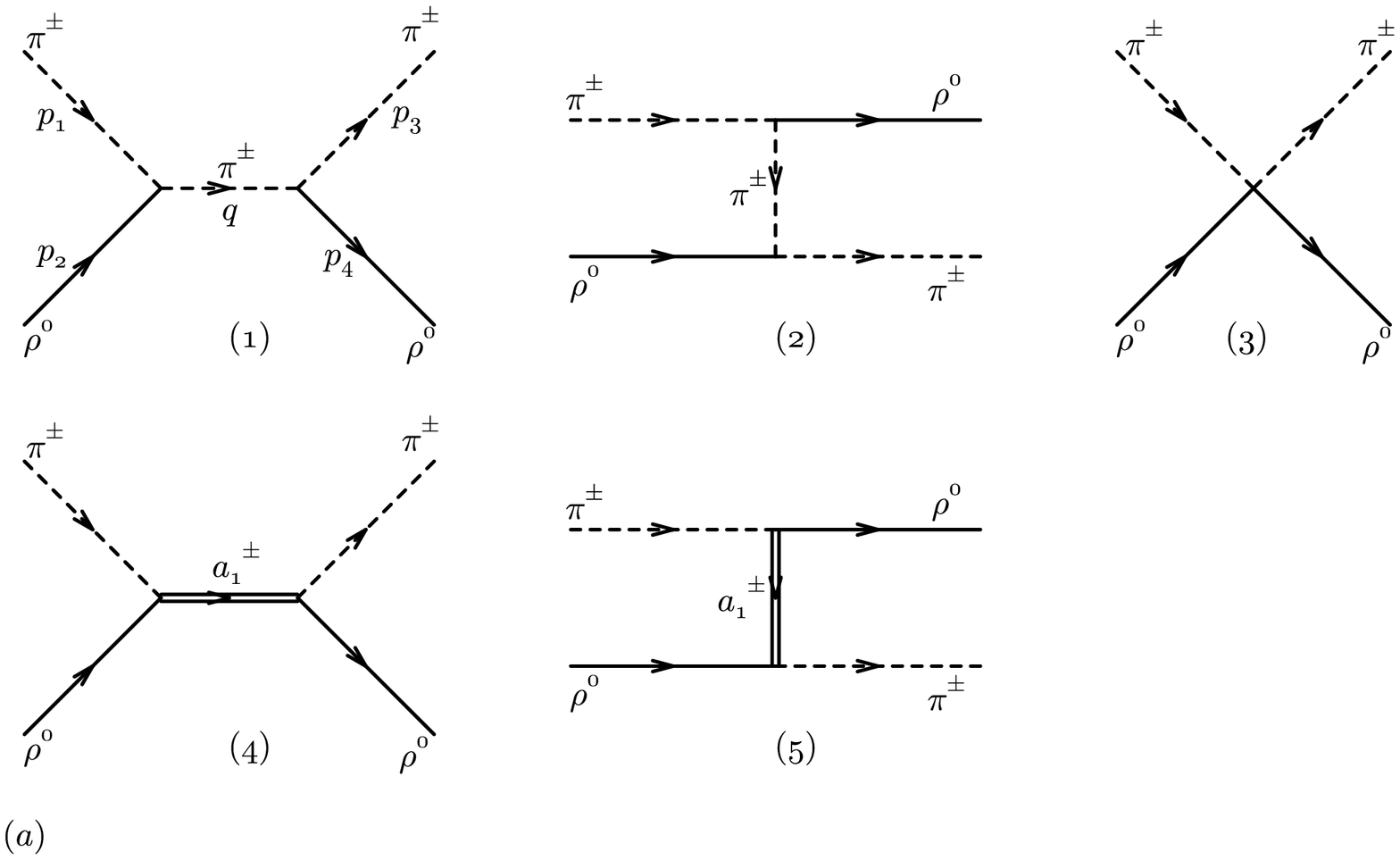,width=80mm,height=80mm,angle=-90}}}
{\makebox{\epsfig{file=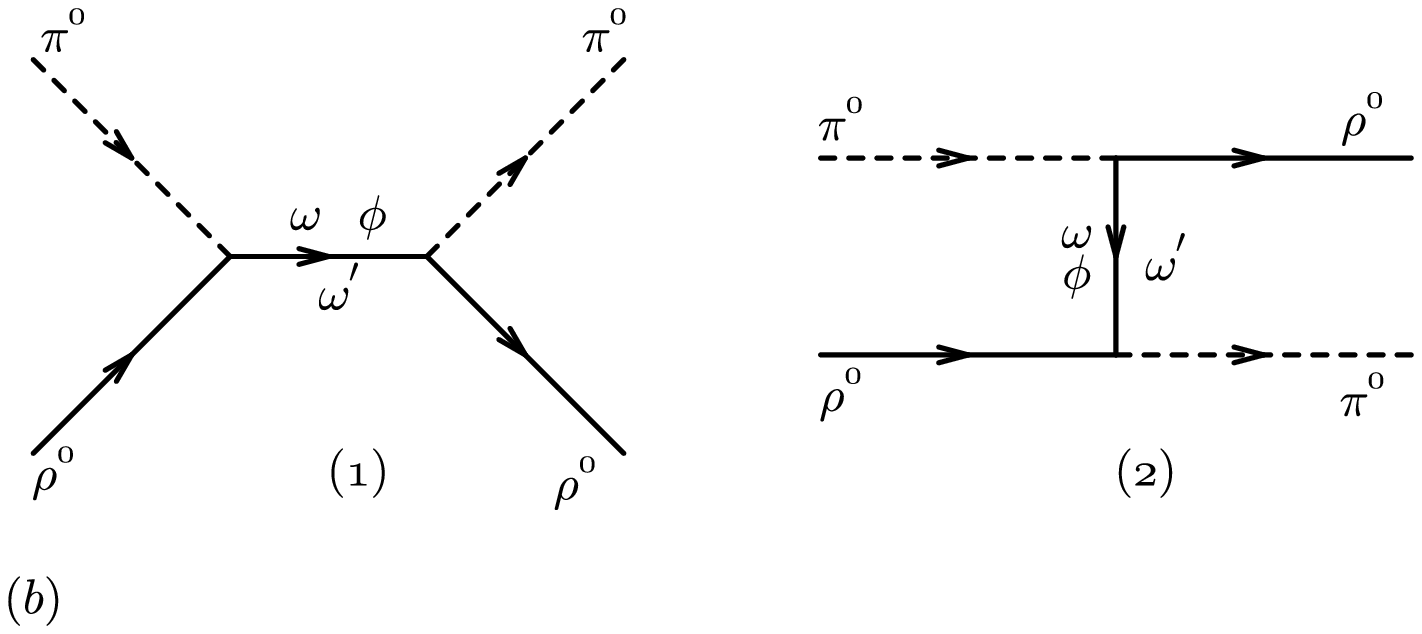,width=80mm,height=80mm,angle=-90}}}
\vspace{-2.cm}
\caption{The Feynman diagrams, (a) for $\rho^0 \pi^{\pm} \to \rho^0
\pi^{\pm}$ reactions and (b) for $\rho^0 \pi^0 \to \rho^0 \pi^0$
reactions.}
\end{figure}

	The pion scattering amplitude on any hadronic target
vanishes at zero pion energy in the target rest frame in the 
zero pion mass limit (Adler's theorem). Following Ref. \cite{elsky} one
can include 
an additional factor 
to the resonance contributions to $\sigma_{\rho\,\pi}$,
\be
\left(\dpf{s-m_{\rho}^2 -m_{\pi}^2}{m_R^2-m_{\rho}^2-m_{\pi}^2} \right)^2 \, ,
\ee
which is normalized to 1 at $s = m_R^2$; for $s > m_R^2$ this factor is 
taken to be one.
The resulting resonant cross section $\sigma_{\rho\,\pi}$
is shown in Fig.~2(a) (dashed curve).

	The cross section for  $\rho\,\pi$ scattering can be calculated by
using an effective Lagrangian for the hadronic interaction  $\rho\,\pi 
\to \rho\,\pi$. The  scattering proceeds through s- and t- channels
respectively, where the $\pi,
\, \rho, \, \omega, \, \phi, \, a_1$ and $\omega'(1420)$ might be
intermediate states. Those processes are constructed from  
 $\pi \pi \rho$, $\pi \rho \omega$ and $\pi \rho a_1$ vertices. 
The effective interaction Lagrangians we use are the following\cite{jhs}:
\begin{mathletters}
\begin{equation}
{\cal L}_{\rho \pi \pi}  =  i g_{\rho \pi \pi} ({\vec \pi} \times
        \partial_{\mu} {\vec \pi} ) \cdot {\vec \rho}\,^{\mu}\, ,
\end{equation}
\begin{eqnarray}
{\cal L}_{\rho \rho\pi \pi} & = & g_{\rho \pi \pi}^2 ({\vec \pi} \times
                {\vec \rho}_{\mu}) \cdot ({\vec \pi}
                        \times {\vec \rho}\,^{\mu}) \, ,
\end{eqnarray}
\begin{eqnarray}
{\cal L}_{\rho \pi \omega} & = & g_{\rho \pi \omega}
        \epsilon^{\mu \nu \sigma \tau} \partial_{\mu} \omega_{\nu}
        \partial_{\sigma}{\vec \rho}_{\tau} \cdot {\vec \pi} \, .
\end{eqnarray}
\end{mathletters}

        The Lagrangian ${\cal L}_{\rho \pi a_1} $ is too long to be
presented
here \cite{gom,gg98} but for the physical $a_1(k) \to 
\rho(q) \pi(p)$ decay, for example,
the vertex function is
\begin{equation}
\Gamma_{\mu \nu}
                =i(f_{A}g_{\mu\nu} + g_{A}q_{\mu}k_{\nu}
                + h_A q_{\mu} q_{\nu}) \, ,
\label{eq:va1}
\end{equation}
where
\begin{eqnarray}
f_{A}&=&\dpf{g}{\sqrt{2}}\left[-\eta_1 q^2
               +(\eta_1-\eta_2)k\cdot q\right] \, , \cr
g_{A}&=& - \dpf{g}{\sqrt{2}}(\eta_1-\eta_2) \, , \cr
h_{A} &=&\dpf{g}{\sqrt{2}} \eta_1 \, .
\end{eqnarray}
The third term in (\ref{eq:va1}) actually does not contribute to the
decay width.  We chose parameters which reproduce the 
measured $a_1 \to \rho \pi$ decay width and D/S ratio 
($g=6.45$, $\eta_1=2.39$ GeV$^{-1}$, $\eta_2 = 1.94$ GeV$^{-1}$).
The coupling constant $g_{\rho \pi \pi} = 6.05$ is determined from a fit
to the free 
$\rho \to \pi \pi$ decay width, $g_{\rho \pi \omega} = 12.40$ GeV$^{-1}$ 
is determined by the $\omega \to \rho \pi \to \pi \gamma$ decay \cite{jhs}. 
The coupling constant $g_{\rho \pi \omega'} =3.55 $ GeV$^{-1}$ is 
determined by the $\omega' \to \rho \pi $ decay. Note that we believe
that it is important for our effective Lagrangians to generate good
phenomenology at energy scales relevant for the applications considered
here. 

	The Feynman diagrams for $\rho \,\pi$ scattering are given in
Fig.~1. The cross section is then calculated as
\be
\sigma_{\rho\,\pi} = \dpf{1}{16\pi 
	\lambda(s,m_{\pi}^2, m_{\rho}^2)} \int_{t_-}^{t_+} {\rm d} t
        {\overline {|{\cal M}|^2}} \, ,
\label{eq:srp2}
\ee
where  
\be
t_{\pm} = m_{\pi}^2 + m_{\rho}^2 -\dpf{1}{2s} [(s + m_{\pi}^2 - m_{\rho}^2)\,
	(s - m_{\pi}^2 + m_{\rho}^2)] \pm \dpf{1}{2s} \lambda(s,m_{\pi}^2,
        m_{\rho}^2) \, .
\ee
For $\rho^0\, \pi^{\pm} \to \rho^0\, \pi^{\pm}$ scattering, 
the squared amplitude is obtained
from the processes shown  
in Fig.~1(a) with the appropriate interference contribution: 
\be
{\overline {|{\cal M}|}^2}= \dpf{1}{(2s_{\pi}+1)(2s_{\rho}+1)} 
	 \,
	|{\cal M}_1 + {\cal M}_2 + {\cal M}_3 + {\cal M}_4 + {\cal M}_5|^2
	\, ,
\label{eq:ampl}
\ee
which  is sum over the spins of the initial states
as well as over those of the final states. 
The $\rho^0\, \pi^{\pm} \to \rho^{\pm}\, \pi^0$ scattering is also
considered, The squared amplitude has the same form as (\ref{eq:ampl}) but 
no contributions from ${\cal M}_2$ and ${\cal M}_5$.
For $\rho^0\,\pi^0$ scattering, Fig.~1(b), 
${\overline {|{\cal M}|}^2}$ can be calculated similarly.

	Due to the substructure of hadrons,
for vertices in the t-channel Feynman diagrams we use form factors 
\be
F_{\alpha} = \dpf{\Lambda^2 - m_{\alpha}^2}{\Lambda^2 - t} \, ,
\ee
where $\alpha$ indicates the 
species of 
the exchange particle. $\Lambda = 1.8$ GeV is taken for all vertices.
The numerical integration for the reaction
$\rho\, \pi \to \pi \to \rho \,\pi$ (t-channel) results in a 
singularity. We regulate it with the effective approach of Peierls\cite{pei}.
This technique offers a pragmatic solution to the possible unitarity
violation at high energies that can result from the use of tree-level
diagrams. A more complete solution would perhaps comprise a K-matrix
multichannel calculation. However, such an endeavor at finite
temperature and finite density also carries ambiguities of its own. 

	After computing the amplitude,  
the $\pi\,\rho$ scattering cross section is obtained from (\ref{eq:srp2}). 
The result is shown in Fig.~2(a) (solid curve).
The resulting $\rho \pi$ cross sections indicate that the difference
between the resonance model and the scattering amplitude method becomes
large at high energies. The resonance model may be valid only at low 
energies. Our effective Lagrangian technique also treats
below-threshold resonances and also receives a contribution from
t-channel diagrams and all their associated quantum interferences. 

\begin{figure}
\vspace{-5.cm}
\hspace{-1.8cm}
{\makebox{\epsfig{file=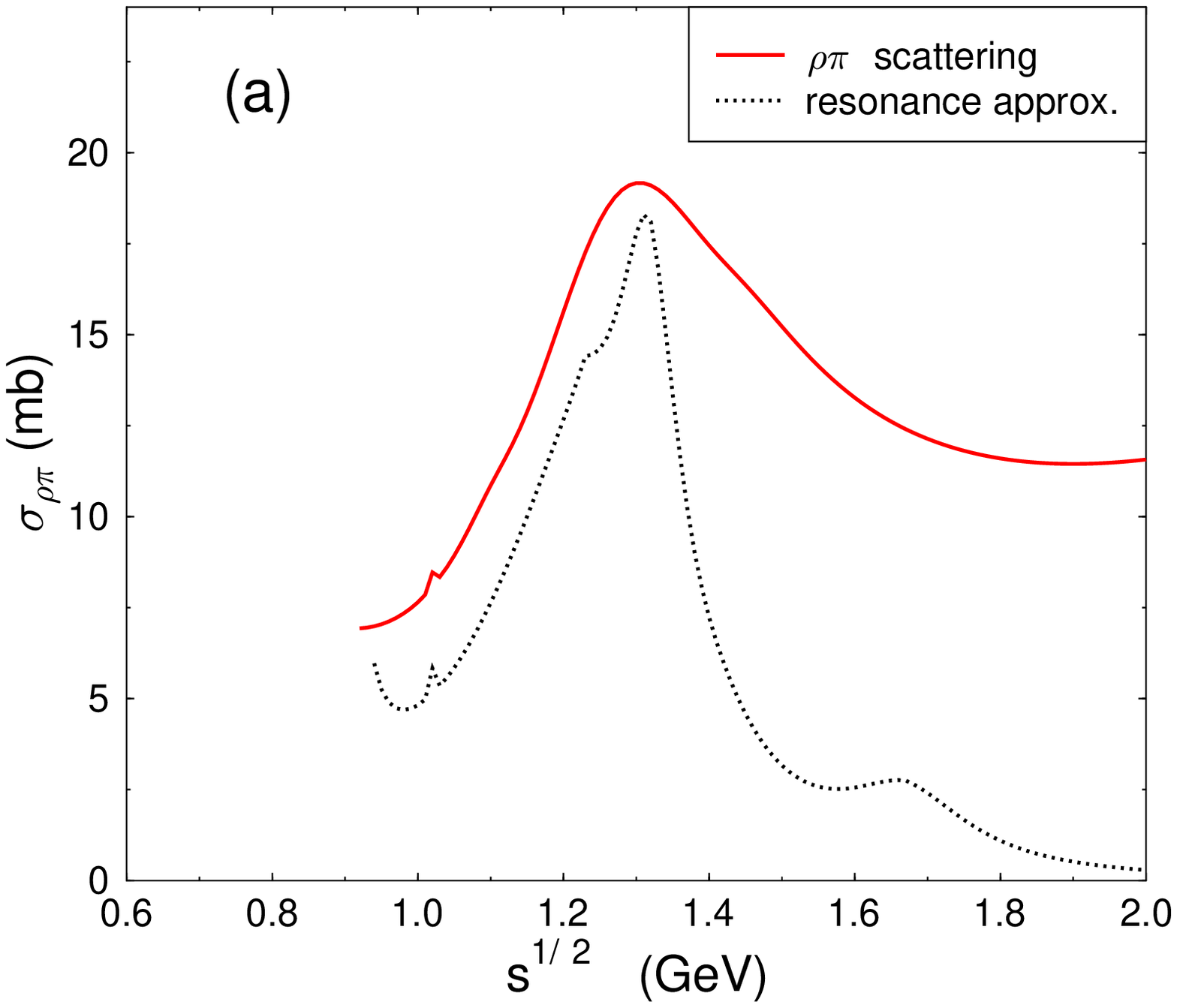,width=9cm,height=7cm}}}
\hspace{-1.6cm}
{\makebox{\epsfig{file=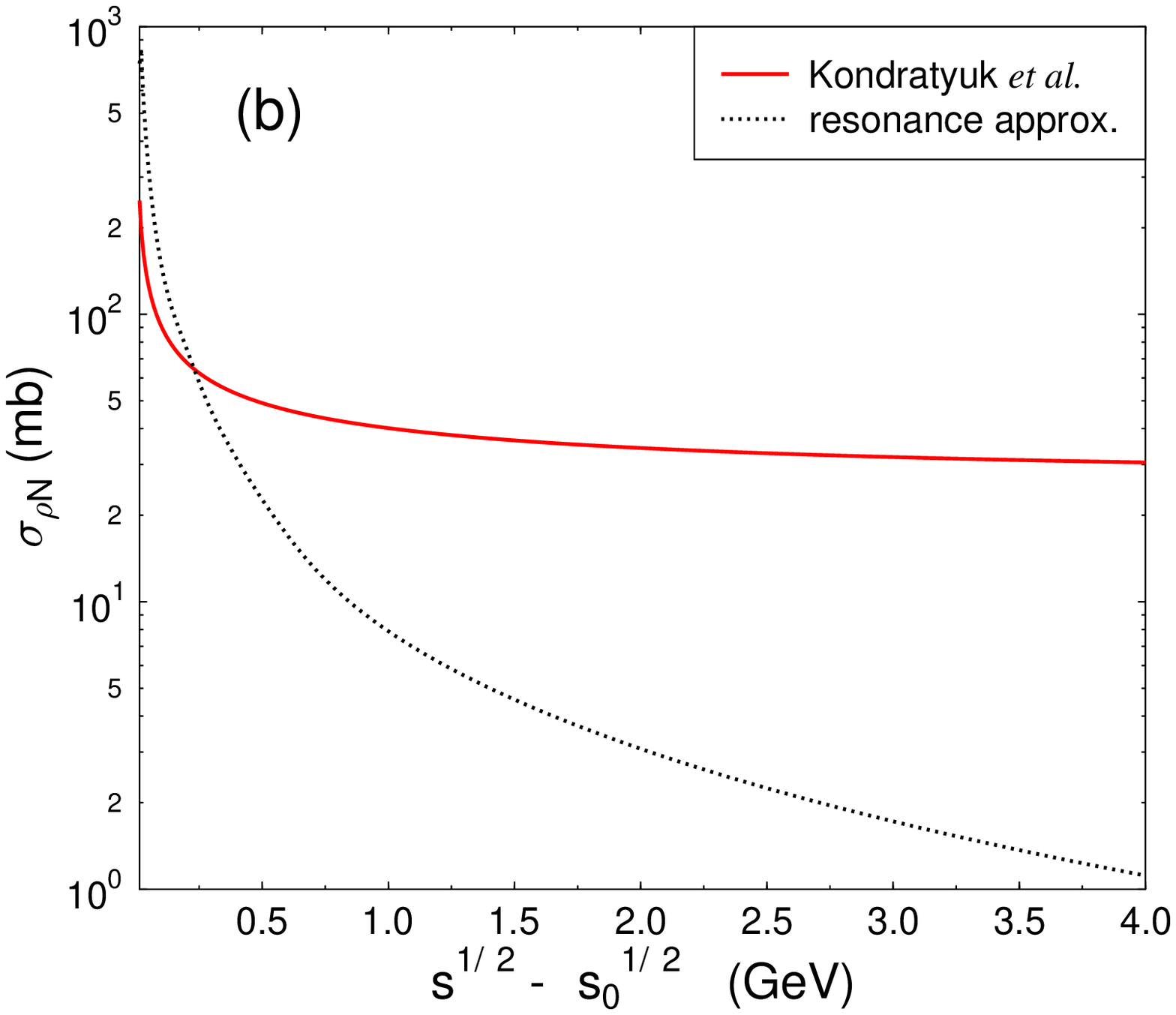,width=9cm,height=7cm}}}
\caption{The cross sections, (a) for $\rho \pi \to \rho \pi$ reactions, and
(b) for $\rho$ N $ \to$ $\rho$ N reactions.}
\end{figure}

\subsection{$\rho$\,N cross section}

	The $\rho$\,N cross section can be calculated within the resonance
model as\cite{elsky,sib98},
\be
\sigma_{\rho\,N} = \dpf{\pi}{6 q^2}\,\sum_R (2J_R +1)\, 
        \dpf{B_R \Gamma_R^2}
	{({\sqrt s}-M_R)^2+\Gamma_R^2/4} \ \ \, .
\label{eq:srn1}
\ee
Here 
\be
 q = \dpf{1}{2{\sqrt s}}\,\lambda^{1/2}(s, m_N^2, m_{\rho}^2)\, ,
\ee
and ${\sqrt s_0} = m_{\rho} + m_N$. 
The summation is performed
over all baryonic resonances with masses above the $\rho$N threshold and
below $2200$ MeV as quoted in 
\cite{PRD}. 
The result of the $\rho$\,N cross section calculated from (\ref{eq:srn1}) 
is shown in Fig.~2(b) (dashed curve).

 	The $\rho$\,N cross section obtained within the resonance model may be
valid only for low energies, while at high energies one
should calculate it from the quark model. Furthermore, it has been 
shown recently
that the resonances below the $\rho$ N threshold, like the N$^*(1520)$
play an important role in the $\rho$-nucleon dynamics \cite{fri98}, so
in the baryonic sector we adopt the $\rho$\,N cross section as determined by
Kondratyuk {\it et al.}
Ref\cite{kondratyuk}, which includes also resonances below the $\rho$N
threshold. The numerical result is shown in Fig.~2(b) (solid curve). 

\section{Broadening the width of the rho at finite 
temperatures and densities}

	The modified properties of hadrons in hadronic matter, {\it e.g.}
the width increase of the particles, are due to their interactions with 
the constituents of the medium. We assume that ultrarelativistic 
heavy ion collisions generate a hot and dense hadron gas of
pions and nucleons and consider the $\rho$ mesons interacting 
with this thermal gas. 
We realize that this is
 a simplification of the complex dynamics which will be
generated in all collision of heavy ions. Our results should
be taken as lower estimates of  the realistic  effects. 

	As discussed in section II, the rho meson width is related to 
the imaginary part of the rho self-energy. The total width of the $\rho$ 
meson is the physical decay width plus the collision width,
\be
\Gamma_{\rho}^{\rm total} = \Gamma_{\rho}^{\rm decay} + 
{\overline \Gamma_{\rho}}^{\rm coll.}  \, .
\ee

	The decay width is related to the imaginary part of the rho
self-energy at the one-loop level. The collision width is related to the
imaginary part of the rho self-energy at the two-loop level. Let us first 
calculate the decay width of the $\rho$ mesons at finite temperature.

\subsection{The decay width of the rho at finite temperature}

	To evaluate the $\rho$ meson self-energy from the $\pi\pi$ loop
we consider the $\rho \pi \pi$ interaction Lagrangian from the previous 
section. Using the Feynman
rules of thermal field dynamics, the temperature dependent rho meson
self-energy is\cite{gsz}
\begin{eqnarray}        
\Pi_{\mu\nu}(q,T) &=&  \dpf{g_{\rho \pi \pi}^2}{(2\pi)^3}
                 \int {\rm d}^4 k
                  \left \{ -2g_{\mu\nu}\delta(k^2-m_{\pi}^2)n_{\pi}(k)
              +(q+2k)_{\mu}(q+2k)_{\nu} \cdot \ \ \right. \cr 
       \ \ \ \ \ \ \ & &\left. \left [ \dpf{\delta(k^2-m_{\pi}^2)}
                      {(k+q)^2-m_{\pi}^2} n_{\pi}(k)
                 +\dpf{\delta((k+q)^2-m_{\pi}^2)}
                  {k^2-m_{\pi}^2} n_{\pi}(k+q) \right ]\, \right \}
                                           \ .
\end{eqnarray}

	Because of the lack of Lorentz invariance in the 
medium\cite{gale91,gsz,kapusta89}, the $\rho$ self-energy at finite
temperature separates into a transverse part, $\Pi_T(q,T)$, and a 
longitudinal part, $\Pi_L(q,T)$. In the rho meson rest frame, 
\be
\Pi(\omega, {\vec q} \to 0, T) = \Pi_L(\omega, {\vec q} \to 0, T) = 
\Pi_T(\omega, {\vec q} \to 0, T) \, 
\ee
thus the real part of the self-energy is,
\be
{\rm Re}\Pi(\omega, {\vec q} \to 0, T) = - \dpf{g_{\rho \pi \pi}^2}{3\pi^2} 
	\int {\rm d} k \dpf{k^2}{E_k} \left (3 - \dpf{4 k^2}{4E_k^2
        - \omega^2}\right )\, n_{\pi}(E_k) \ , 
\ee
where $E_k = {\sqrt {k^2 + m_{\pi}^2}}$. The real self-energy will lead 
to a mass shift of the $\rho$ at finite temperature, which is determined by 
the solution of 
\be
\omega^2 - m_{\rho}^2 + {\rm Re}\Pi(\omega, {\vec q} \to 0, T) = 0 \, .
\ee
It is known from previous calculations that this mass shift is small
\cite{gale91}. 

	The imaginary part of the self-energy can be obtained by using
cutting rules \cite{weldon83},
\be
{\rm Im} \Pi(\omega, T) = \dpf{g_{\rho \pi \pi}^2}{6 \pi} 
	\int {\rm d} k \dpf{k^4}{E_k} \left (2 n_{\pi}(E_k)+1 \right )\ \
	\left [\delta(\omega + 2 E_k) - \delta(\omega -  2 E_k)
	\right ] \ .
\ee
	From ${\rm Im} \Pi(\omega, T)$, the corresponding width of the 
$\rho \to \pi \pi$  decay follows to be
\be
\Gamma_{\rho}^{\rm decay}(T) = \dpf{g_{\rho \pi \pi}^2}{48 \pi \omega^2}\, 
	(\omega^2 - 4 m_{\pi}^2)^{3/2} \left [2 n_{\pi}(\dpf{\omega}{2}) + 1
	\right ] \ .
\ee

	This decay width varies with the temperature as shown in Fig.~3. 
Obviously the decay width increases with temperature as well as with 
the pion chemical potential. For $T=150$ MeV, when $\mu_{\pi} = 0$, 
$\Delta \Gamma_{\rho} = \Gamma_{\rho}^{\rm decay}(T) - 
\Gamma_{\rho}^{\rm decay}(0) = 25$ MeV, while for $\mu_{\pi} = 135 $ MeV, 
$\Delta \Gamma_{\rho} = 70 $ MeV. 
\begin{figure}[hbt]
\begin{minipage}[t]{70mm}
\vspace{-5.5cm}
\hspace{-0.4cm}
{\centerline{\epsfig{file=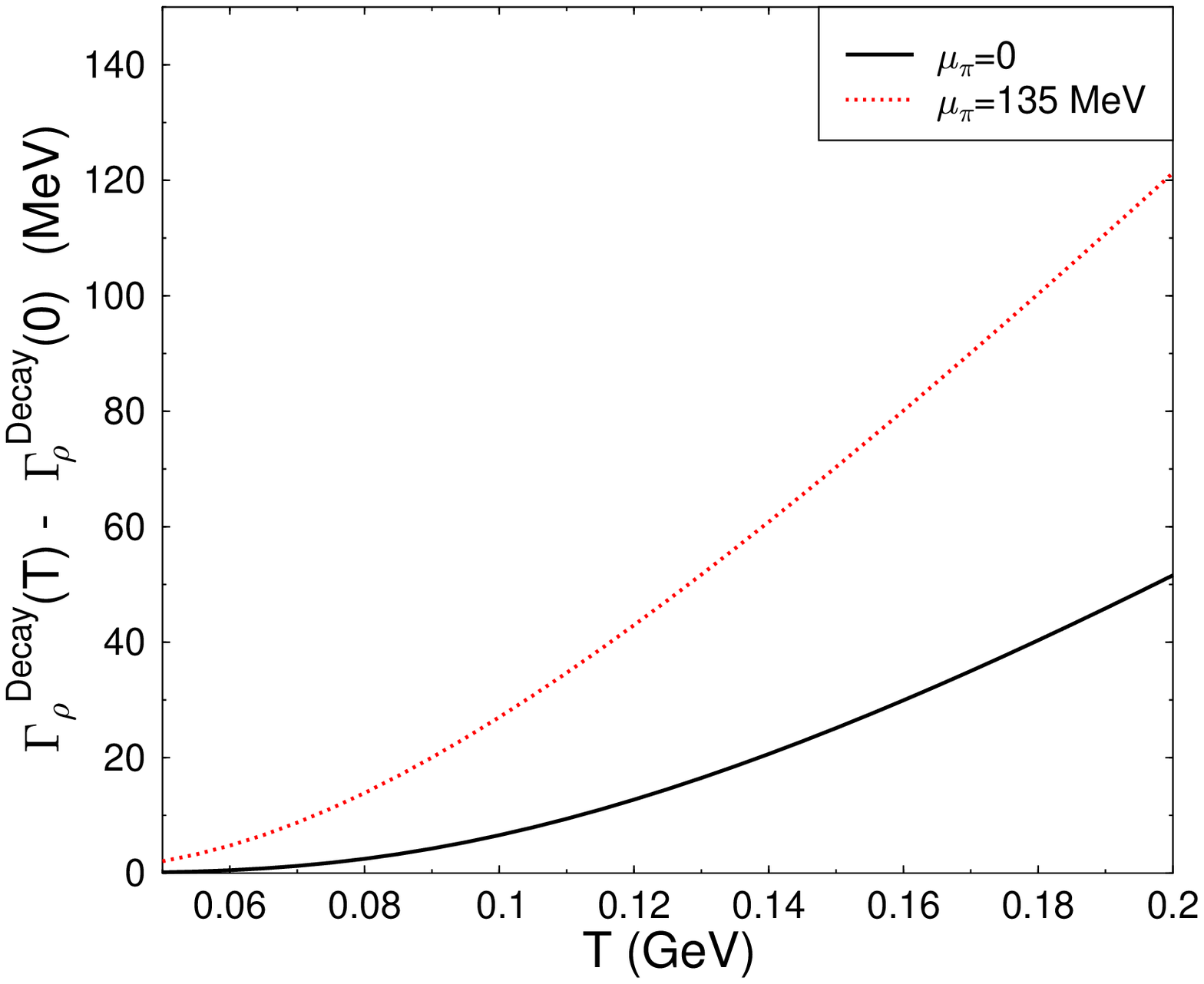,width=85mm,height=55mm}}}
\caption{The $\rho \to \pi \pi$ decay width as a function of temperature $T$
at two fixed values of the pion chemical potential.}
\label{fig:dcw}
\end{minipage}
\vspace{0.5cm}
\hfill
\begin{minipage}[t]{70mm}
\vspace{-5.5cm}
\hspace{-0.6cm}
{\centerline{\epsfig{file=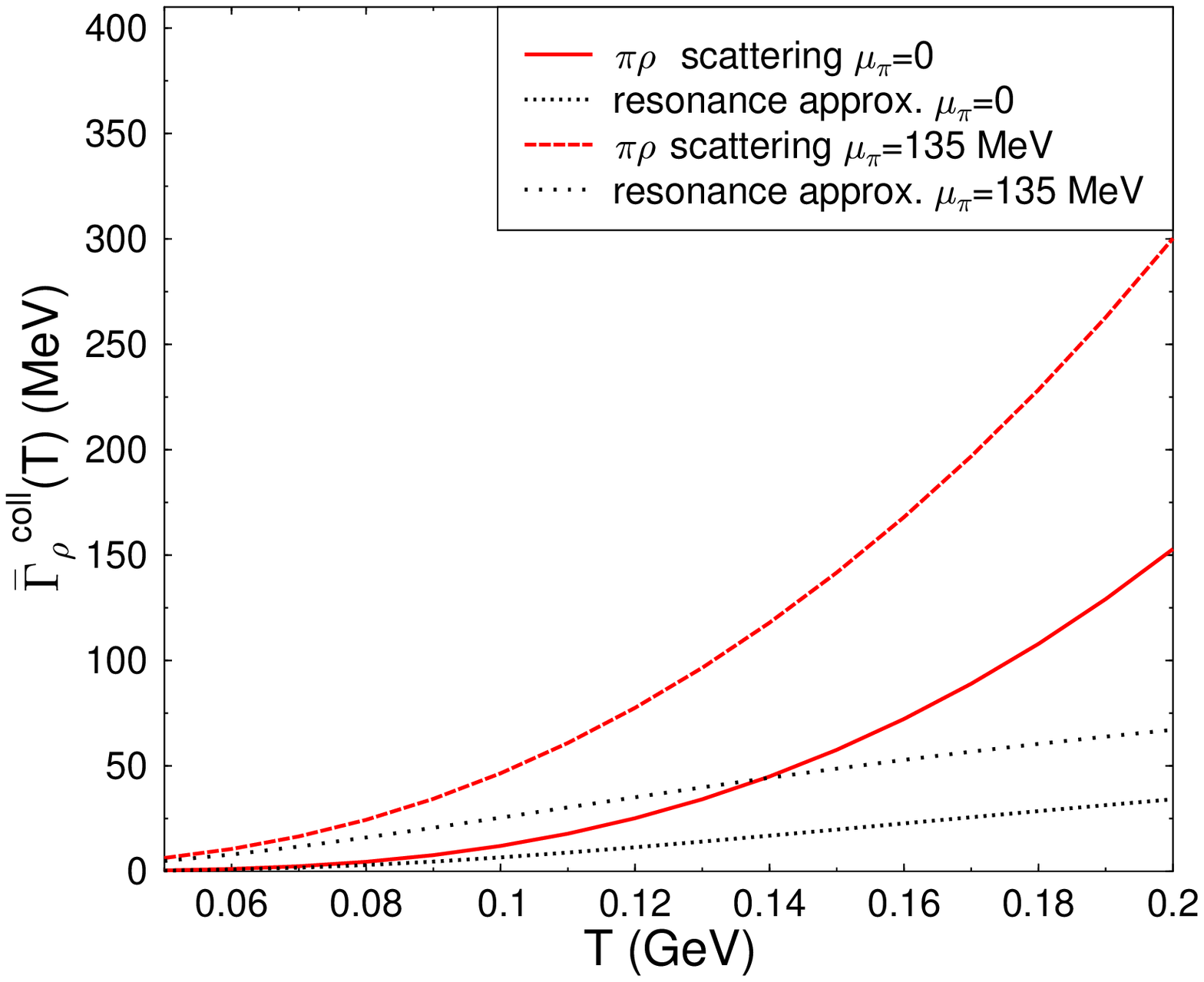,width=85mm,height=55mm}}}
\caption{The $\rho$ collision rate $vs.$ the temperature due to
$\rho \, \pi$ interaction in two different approaches.}
\end{minipage}
\vspace{0.5cm}
\end{figure}

\subsection{Collision rate of rho at finite temperature and density}
 
        From (\ref{eq:aver}) and with the cross sections computed
in the previous section, 
the collision rates of the $\rho$ mesons can be calculated numerically.

	First the rho collision rate is calculated due to $\rho \pi$
reactions in the finite temperature meson medium. The contributions from 
$\rho^0 \pi^{\pm} \to \rho^0 \pi^{\pm}$ , $\rho^0 \pi^{\pm} \to 
\rho^{\pm} \pi^0$ and $\rho^0 \pi^0 \to \rho^0 \pi^0$ 
processes are taken into account. Fig.~4  shows the 
the collision rates as calculated 
from the scattering amplitude calculation and from 
the Breit-Wigner resonance approximation at two values of the 
pion chemical
potential, $\mu_{\pi} = 0$ and $\mu_{\pi} = 135 $ MeV. 
For the same temperature and chemical potentials and using the $\rho
\pi$ cross section from the scattering amplitude method, 
one can get a larger
collision rate of $\rho$ mesons.   

	The collision rate from $\rho$ N reactions is plotted in Fig.~5. 
In Fig.~5(a), we use  $T= 100, \ \ 150 $ MeV and show the
rate as a function of the nucleon density ratio. In Fig.~5(b), we show the 
rate as a function of the temperature for two values of the 
nucleon chemical potential,
$\mu_N = 0.4, \ \ 0.6$ GeV. From Fig.~5(a), we can see the collision
rate increasing with the nucleon density for a given temperature, but
decreasing with temperature for a fixed nucleon density. The latter 
decrease is due to the fact that 
the nucleon chemical potential decreases with
temperature at a fixed nucleon density. 
Again we find that for given temperature
and chemical potential the pure resonance contribution
yields a lower collision rate than the resonance + quark
model.

\begin{figure}
\vspace{-5.cm}
\hspace{-1.8cm}
{\makebox{\epsfig{file=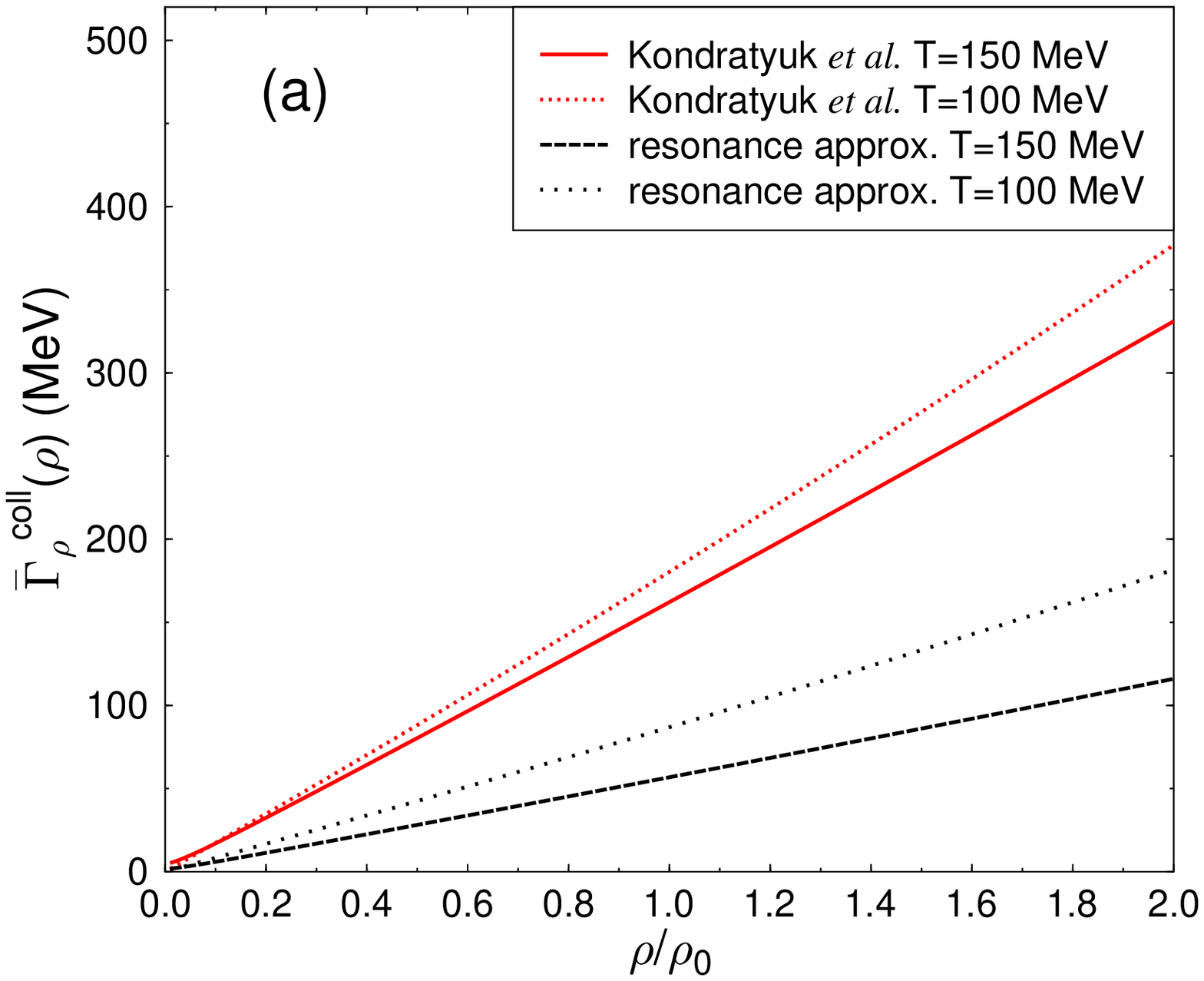,width=9cm,height=12.5cm}}}
\hspace{-1.6cm}
{\makebox{\epsfig{file=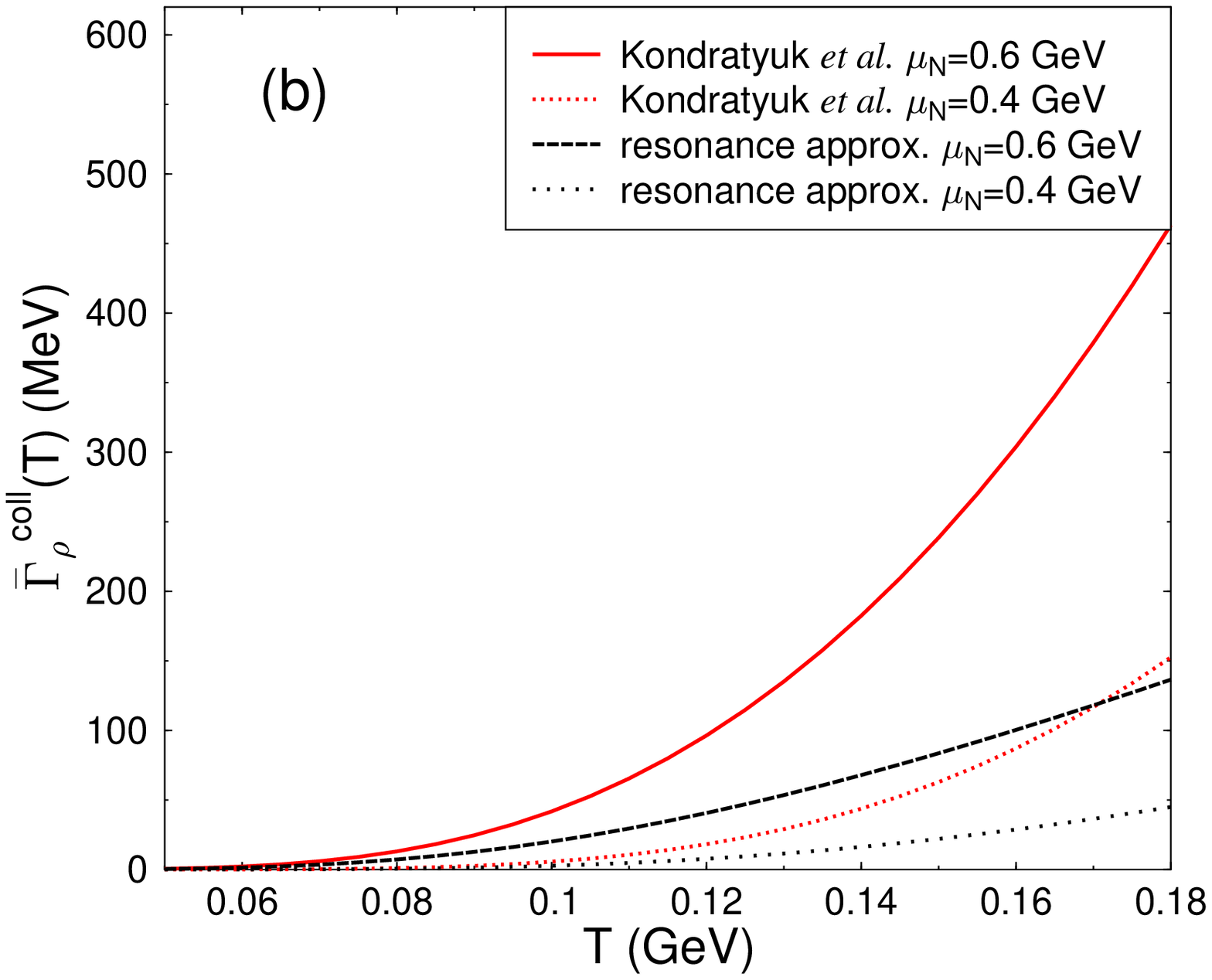,width=9cm,height=12.5cm}}}
\caption{The collision rate of the $\rho$ due to $\rho$ N interactions,
$vs.$ (a) the nucleon density for vary temperature.
(b) the temperature for vary baryon chemical potential.}
\end{figure}

        We employ the resonance model modified to include the 
resonances below the
$\rho$N threshold, such as N$^*(1520)$, as described previously. The 
total $\rho$N cross section can
be evaluated as a function of the invariant collision energy ${\sqrt s}$ and
the invariant mass of the $\rho$ meson $M_{\rho}$. In this case, the
collision width has to be integrated over the spectral function of
the $\rho$ meson as 
\be
{\overline \Gamma}_{\rho {\rm N}} =\dpf{{\cal N}_1 {\cal N}_2}{\rho_1}\,
        \int_{s_0'}^{\infty} {\rm d} s \dpf{T}{2(2\pi)^4 {\sqrt s}}\,
        \int_{2m_{\pi}}^{{\sqrt s}-m_N} {\rm d} M_{\rho} \, M_{\rho}
	A(M_{\rho}) \, 
	\lambda(s, M_{\rho}^2, m_N^2)\, {\tilde K}\,
        \sigma_{\rho {\rm N}}(s, M_{\rho}) \, ,
\label{eq:massint}
\ee
here $s_0'=(2m_{\pi}+m_N)^2$. $A(M_{\rho})$ is the spectral function of
the $\rho$ meson in free space taken as 
\be
A(M_{\rho}) = \dpf{1}{\pi} \dpf{m_{\rho} \Gamma_{\rho}(M_{\rho})}
	   {(M_{\rho}^2-m_{\rho}^2)^2 + m_{\rho}^2 \Gamma_{\rho}^2} \, ,
\ee
where $m_{\rho} = 770$ MeV, $\Gamma_{\rho}(M_{\rho})$ is the mass-dependent
width of the $\rho$ meson\cite{kondratyuk}. For the same process of  
$\rho$N scattering, the collision
width of the $\rho$ meson in this calculation is about 10\% larger than 
that from the on mass-shell $\rho$ meson calculations.  

	In summary, our results indicate that at high temperature 
and low nucleon
density, the collision rate is dominated by $\rho$ + meson reactions, 
while at
low temperature and high nucleon density it is dominated by 
$\rho$ + nucleon reactions. 
For example, when $T=100$ MeV, $\mu_{\pi} =0$, $\rho_N = \rho_0$,
${\overline \Gamma_{\rho \pi}}^{\rm coll.} = 12$ MeV, while 
${\overline \Gamma_{\rho N}}^{\rm coll.} = 180$MeV.

For conditions typical of high energy heavy-ion collisions 
\cite{rapp,sorge}, $T=150$ MeV, $\mu_{\pi} =0$ and $\mu_N = 0.4$ GeV,
($\rho_N = 0.39 \rho_0$ ) the total width of rho mesons is
\bea
\Gamma^{\rm total} &=& \Gamma^{\rm decay} + {\overline 
                     \Gamma_{\rho \pi}}^{\rm coll.} +
		     {\overline \Gamma_{\rho N}}^{\rm coll.} \, \cr 
		&\approx& 176 + 58 + 63 = 297\, {\rm MeV} \ \ . \nonumber
\eea
This dramatic increase of the $\rho$ width is most certainly important. 
\begin{figure}
\vspace{-5.0cm}
\hspace{-1.8cm}
{\makebox{\epsfig{file=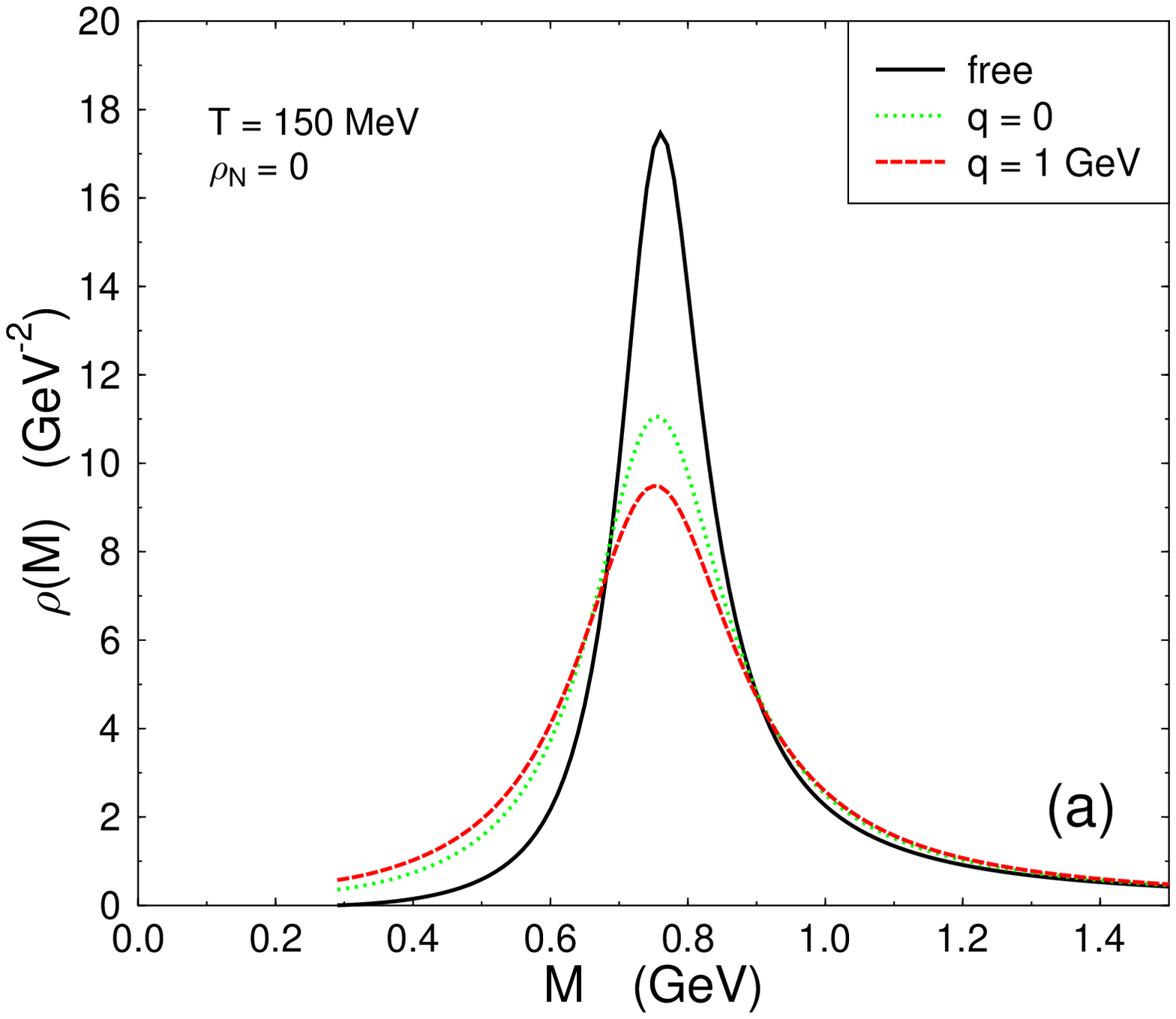,width=9cm,height=12.5cm}}}
\hspace{-1.6cm}
{\makebox{\epsfig{file=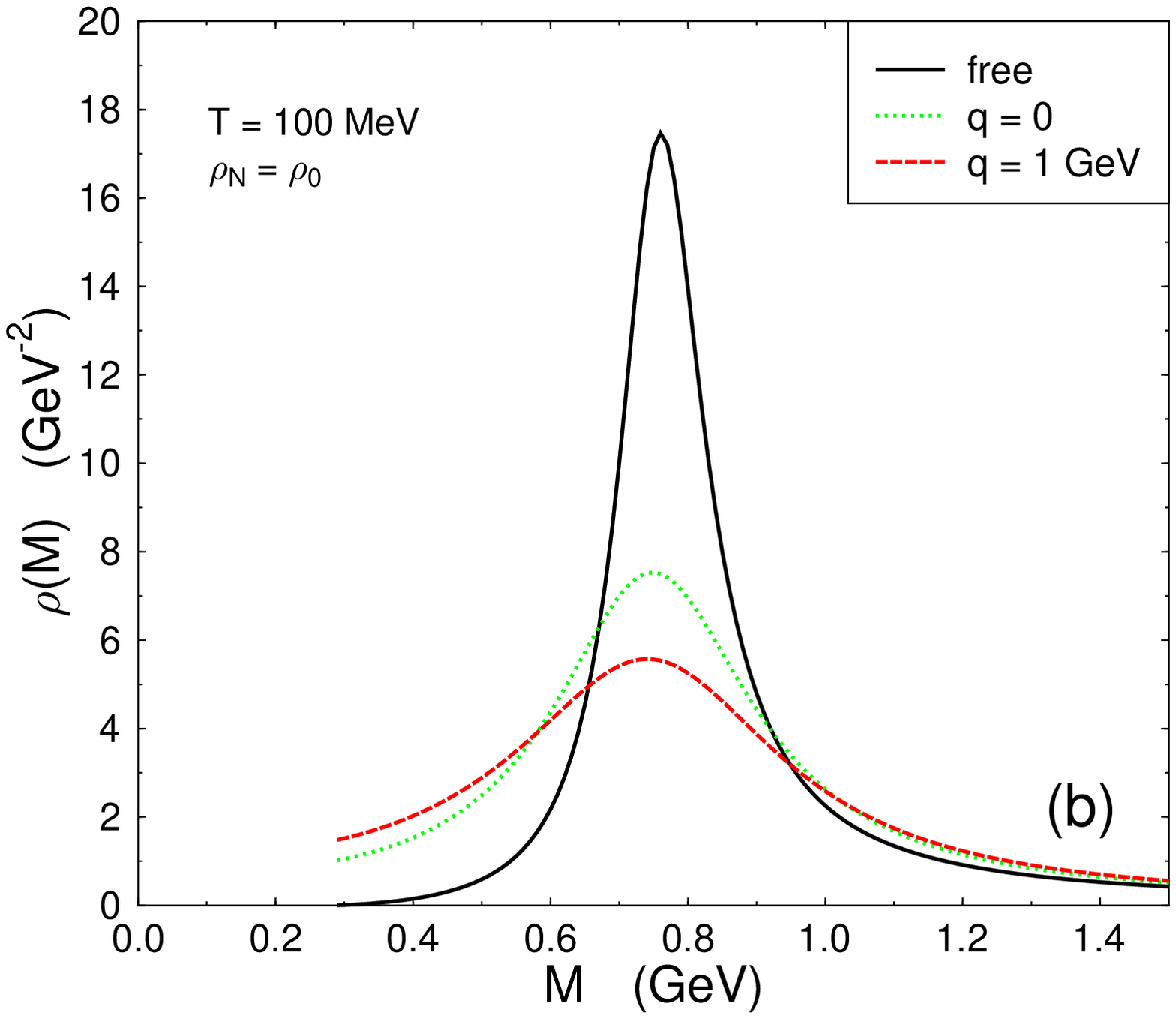,width=9cm,height=12.5cm}}}
\caption{The spectral function of the $\rho$ $vs.$ the invariant mass M
of the $\rho$ for fixed values of momentum as indicated.
(a) In pure pion gas. (b) In the matter of pions and nucleons.}
\end{figure}

The spectral function of rho meson is given by
\be
\rho(M) = - \dpf{2 {\rm Im} \Pi}{(\omega^2-q^2-m_{\rho}^2 -{\rm Re} \Pi)^2 + 
	({\rm Im} \Pi)^2} \, ,
\ee
where $q$ is the three-momentum of the $\rho$ meson. 
The vacuum part of the self-energy $\Pi$, which is given in
\cite{elkap},
can only depend on the invariant mass $M ={\sqrt {\omega^2-q^2}}$.
We plot the $\rho$ meson spectral function in fig.~6(a) for a pure pion
gas at a temperature $T=150$ MeV and in fig.~6(b) for the matter of 
pions and nucleons at $T=100$ MeV and nucleon density 
$\rho_N = \rho_0 = 0.16$ fm$^{-3}$.  We find a
broadening of the $\rho$ spectral function and a sizable
suppression of the peak strength. Because the cross sections
from the collision rate are evaluated on the $\rho$ meson
mass shell, the imaginary self energy of $\rho$ which comes from the
collisions does not vary with $M$. Therefore the spectral
function doesn't vanish when the two-pion threshold
is reached. 

\section{Summary}	

   	Based on the idea that in thermal systems a particle can be captured
and can also be produced by the thermal background, and the fact that 
the collision rate
is the difference of the capture and the production rate, we have calculated 
the $\rho$ decay width at finite temperature and
the collision rates of the $\rho$ due to the $\rho\, \pi$ scattering 
and $\rho$N scattering in hot and dense hadronic matter. In 
high temperature and/or high density hadronic matter, the collision 
rate is much larger than the decay width correction  
due to the one-loop
self-energy modifications. For the collision rates, the contribution from 
$\rho\,\pi$ collisions is the most important one in the high temperature
pion gas, while  at low temperatures and high density nuclear matter 
the $\rho$N contribution is more important.
 
       The collision rate of $\rho$ mesons with pions 
 uses an effective 
 Lagrangian for the $a_1 \rho \pi$ interaction which is tuned to hadronic
 phenomenology \cite{gom,gg98}.
 Note that in $\rho \pi \to \rho \pi $
 scatterings, the $a_1$ intermediate state gives the largest contribution
 to the cross section and to the collision rate. 
 Our result for $\rho$N
 scattering is in qualitative agreement with the width increase found in
 \cite{elkap}. 

	We have shown that the width of the $\rho$ is increased 
drastically
through its interaction with pions and nucleons present in typical 
heavy-ion collisions. 
It will be even more increased if finite 
pion chemical potentials are introduced or if other scattering partners are
considered.
This result will be important not only for the 
dilepton spectra, but also for the general dynamics of heavy ion 
collisions.
We believe that our results
adds to the consensus building in this direction.

\acknowledgments
	S.\ Gao thanks the Alexander von Humboldt-Stiftung for financial
support.
C. G. is happy to
acknowledge useful discussions with R. Rapp. 
	This work was supported  by the Graduiertenkolleg Theoretische 
und Experimentelle Schwer\-ionen\-physik, BMBF, DFG, GSI, by the
Natural Sciences and Engineering Research Council of Canada, and 
by the Fonds FCAR of the Quebec Government.

\end{document}